\documentclass[%
reprint,
superscriptaddress,
amsmath,amssymb,
aps,
prb,
citeautoscript
]{revtex4-1}
\usepackage[T1]{fontenc}
\usepackage{upgreek}
\usepackage{amsmath}
\usepackage{gensymb}
\usepackage{braket}
\usepackage{multirow}
\usepackage{graphicx}% Include figure files
\usepackage{dcolumn}% Align table columns on decimal point
\usepackage{bm}% bold math
\usepackage{float}
\usepackage{color,soul}
\usepackage{textcomp}
\usepackage{hyperref}% add hypertext capabilities
\usepackage{lmodern}
\graphicspath{ {./} }
\begin{document}
	
	\preprint{APS/123-QED}
	
	\title{Electronic structure and magneto-optical properties of silicon-nitrogen-vacancy complexes in diamond}
	
	\author{Marcin Roland Zem{\l}a}
    \email[]{Marcin.Zemla@pw.edu.pl}
    \affiliation{Materials Design Division, Faculty of Materials Science and Engineering, Warsaw University of Technology, Wo{\l}oska 141, 02-507 Warsaw, Poland}
    \affiliation{Institute of Theoretical Physics, Faculty of Physics, University of Warsaw, Pasteura 5, 02-093 Warsaw, Poland}
    \author{Kamil Czelej}
    \email[]{Kamil.Czelej@fuw.edu.pl}
    \affiliation{Institute of Theoretical Physics, Faculty of Physics, University of Warsaw, Pasteura 5, 02-093 Warsaw, Poland}
    \author{Paulina Kami{\'n}ska}
    \affiliation{Materials Design Division, Faculty of Materials Science and Engineering, Warsaw University of Technology, Wo{\l}oska 141, 02-507 Warsaw, Poland}
    \author{Chris G. Van de Walle}
    \affiliation{Materials Department, University of California, Santa Barbara, California 93106-5050, USA}
    \author{Jacek A. Majewski}
    \affiliation{Institute of Theoretical Physics, Faculty of Physics, University of Warsaw, Pasteura 5, 02-093 Warsaw, Poland}

    \date{\today}
	
	\begin{abstract}
    	The silicon-vacancy (SiV) and nitrogen-vacancy (NV) centers in diamond are commonly regarded as prototypical defects for solid-state quantum information processing. Here we show that when silicon and nitrogen are simultaneously introduced into the diamond lattice these defects can strongly interact and form larger complexes. Nitrogen atoms strongly bind to Si and SiV centers and complex formation can occur. Using a combination of hybrid density functional theory (DFT) and group theory, we analyze the electronic structure and provide various useful physical properties, such as hyperfine structure, quasi-local vibrational modes, and zero-phonon line, to enable experimental identification of these complexes. We demonstrate that the presence of substitutional silicon adjacent to nitrogen significantly shifts the donor level toward the conduction band, resulting in an activation energy for the SiN center that is comparable to phosphorus. We also find that the neutral SiNV center is of particular interest due to its photon emission at $\sim$1530 nm, which falls within the C band of telecom wavelengths, and its paramagnetic nature. In addition, the optical transition associated with the SiNV$^0$ color center exhibits very small electron--phonon coupling (Huang--Rhys factor~=~0.78) resulting in high quantum efficiency (Debye-Waller factor = 46\%) for single-photon emission. These features render this new center very attractive for potential application in scalable quantum telecommunication networks.
	
	\end{abstract}
	\maketitle

	\section{INTRODUCTION}
    	Paramagnetic point defects embedded into the diamond host are arguably the most thoroughly investigated solid-state systems for single-photon emitters and quantum information processing applications \cite{Weber2010,Czelej_JMCC, Kurtsiefer2000, Aharonovich2009, Simpson2009}. Amongst a variety of well-known defect complexes in diamond, the negatively charged NV \cite{Gruber2012, Deak2014, Dutt1312} and SiV \cite{Goss1996, Thiering2018PRX, Neu_2011, Gali2013PRB, Hepp2014, Muller2014, Rogers2014} centers have attracted a great deal of attention and have became the prototype solid-state quantum bits (qubits).

        The NV center consist of a nitrogen atom adjacent to a carbon vacancy and has an total spin $S$ = 1 ground state with millisecond coherence time at room temperature, in the case of a $^{13}$C-enriched diamond host. The electronic structure of NV$^{-}$ centers has been precisely described based on group theory considerations \cite{Thiering2018PRB}, advanced hybrid density functional theory (DFT) \cite{Thiering2018PRB, Gali2009PRL}, and fully correlated configuration interaction calculations\cite{Delaney2010}. In the C$_{3v}$ crystal field the ground state of NV$^{-}$ is a spin triplet of $^3$A$_2$ symmetry, split by 2.88 GHz into two sublevels due to the dipolar electron-spin--electron-spin interaction.
        The first spin-conserving excited state of NV$^{-}$ is the triplet $^3$E.
        Group theory analysis \cite{Doherty_2011,Maze_2011} combined with absorption \cite{Kehayias2013} and luminescence \cite{Rogers_2008} experiments revealed two shelving singlet states $^1$A$_1$ and $^1$E between the $^3$A$_2$ ground-state and $^3$E excited-state triplet, separated by 1.19 eV.
        An essential property of NV centers is the correlation between the electron spin resonance and the spin-selective fluorescence intensity that can be used for quantum bit initialization and readout protocols. Unfortunately, the optical properties of this center are strongly influenced by electron-phonon coupling of its excited states. This causes a broad emission ($\sim$100 nm), of which only 4\% is associated with the zero-phonon line (ZPL).

        \parskip = 0pt The split-vacancy SiV center belongs to a very attractive class of complexes between group-IV impurities and vacancy defects in diamond. The presence of an inversion center in the D$_{3d}$ point group, to which the SiV center belongs, leads to an almost complete absence of a Stark shift in the optical signals, an essential feature for robust indistinguishable single-photon emitters. \cite{Thiering2019CM} The negatively charged SiV$^{-}$ exhibits a large Debye-Waller (DW) factor of 0.7, a favorable characteristic for its utility in quantum communication and sensor applications.\cite{Sipahigil847,Kucsko2013} Nevertheless, a main drawback of the SiV$^{-}$ center is its very short spin coherence time ($T_2$), limited by an orbital relaxation time ($T_1$) of $\sim$40 ns at T = 5 K \cite{Becker2016}, which is caused by phonon dephasing mediated by the dynamic Jahn-Teller effect on the orbital doublet.\cite{Jahnke_2015} This limitation, however, has been overcome by proper Fermi-level engineering, enabling the stabilization of neutral SiV in a triplet $S$ = 1 spin state. In fact, it has been recently demonstrated that SiV$^0$ exhibits a spin coherence time of nearly one second at 20 K and a near-infrared optical transition at 946 nm.\cite{Rose2019}

        Although very extensive research on both NV and SiV centers in diamond has been performed, the manner in which nitrogen, silicon, and vacancies interact with one another when they are simultaneously introduced into the crystal remains to be clarified. Very recently, Breeze \emph{et al.} for the first time detected spectroscopic signatures of a new Si- and N- related center in diamond.\cite{breeze} Based on experimental measurements combined with DFT calculations, they were able to unambiguously assign this new center to the neutral SiNV complex. Wassell \emph{et al.}, on the other hand, speculated that either Si$_x$N$_y$ or Si$_x$N$_y$V complexes, where $x$ and $y$ are small integers, might be responsible for the 499 nm line observed in the UV-Vis luminescence spectrum of synthetic CVD diamond. \cite{Wassell2018} To the best of our knowledge, various complexes of Si--N--V have not been thoroughly investigated so far and their electronic structure and magneto-optical properties remain elusive.

        Here, we apply an advanced spin-polarized hybrid density functional theory (SP-DFT) method to investigate the energetics, electronic structure, and magneto-optical properties of yet unidentified silicon, nitrogen and vacancy related complexes in diamond.
        First, we analyze the stability of SiN, SiNV, and SiN$_2$V centers as a function of the charge state and the position of the Fermi level in the band gap.
        We find that a strong thermodynamic driving force exists for complexing of silicon and nitrogen, and for the formation of complexes with vacancies.
        In the next step, we extensively discuss the electronic structure of the selected complexes by using the elements of group theory and the Kohn-Sham eigenvalue spectra calculated with the hybrid functional.
        The SiN dimer is found to have a donor level at 0.57 eV below the conduction-band minimum (CBM).  This value is comparable to that of substitutional phosphorus, which is regarded as the shallowest donor in diamond.  The SiN center could therefore be very attractive for generating $n$-type conductivity in diamond.
        
        To facilitate experimental identification of the silicon--nitrogen--vacancy centers in diamond, we calculate their electron paramagnetic resonance (EPR) signals, quasi-local vibrational modes, and optical fingerprints.
        Standing out among our detailed results is the finding that the SiNV$^0$ center is an optically active color center.
       Our calculated ZPL energy for the a' $\rightarrow$ a'' transition is at $\sim$1530 nm, which falls within the C band of telecom wavelengths.
       To analyze the quality of quantum emission from the SiNV$^0$ center we also investigate the Huang-Rhys (HR) factor, and finds a value (0.78) indicating a high quantum efficiency (Debye-Waller factor = 46\%). This new center is, therefore, very attractive for potential applications as a bright single-photon emitter in scalable quantum telecommunication networks.
        	
	\section{THEORY AND COMPUTATIONAL DETAILS}
    \label{sec:theory}
    
    \subsection{Electronic and atomic structure}

    	The spin-polarised electronic structure calculations have been carried out using SP-DFT and the projector augmented wave method (PAW) \cite{PAW1_Bloch_PRB, PAW2_Kresse_PRB} pseudopotentials, as implemented in the Vienna Ab initio Simulation Package (VASP). We employed the screened, range separated, non-local hybrid functional of Heyd, Scuseria and Ernzerhof,\cite{HSE1_JCP, HSE2_JCP} (HSE06) with standard values of the $\alpha=0.25$ and $\omega=0.2$ parameters to calculate the ground-state charge and spin densities of the system. It has been shown that HSE06 satisfies the generalized Koopmans' theorem in group-IV semiconductors \cite{Lany_PhysRevB.80.085202}. HSE06 in diamond turned out to be nearly free of the electron self-interaction error, due to the error compensation between the Hartree-Fock and generalized-gradient-approximation (GGA) exchange, and thus capable of providing defect levels and defect-related electronic transitions close to experimental values \cite{Czelej_MRSAdv2017}.
    	
    	Convergence parameters and equilibrium lattice constants   were determined from bulk calculations on a primitive cell.	
    	To ensure convergence of the charge density, the Brillouin zone (BZ) sampling with $8\times{8}\times{8}$ Monkhorst-Pack mesh and the plane-wave cutoff energy of 520~eV were applied. The resulting relaxed lattice parameter $a_{HSE}=3.594$~{\AA} and indirect band gap $E_g=5.32$~eV agree well with the experimental values, $a=3.567$~{\AA} and $E_g=5.48$~eV \cite{Yamanaka_PhysRevB.49.9341}.
    	
        In order to minimize finite-size effects we selected a large cubic 512-atom supercell. This enables accurate sampling of the first Brillouin zone (BZ) using the $\Gamma$-point, allowing, in turn, inspection of Kohn-Sham wave functions with correct symmetry and degeneracy. Previous studies have shown that this setup reproduces known experimental values in the case of point defects in diamond \cite{Czelej_MRSAdv2017, Czelej_JMCC, Czelej_MRSAdv2016, Czelej_Diamond} and silicon \cite{Spiewak_PhysRevB.88.195204}. Defects in the supercell were allowed to relax in constant volume until the Hellmann-Feynman forces acting on atom were smaller than 0.01 eV/\AA. The same plane-wave cutoff energy of 520~eV was applied for the defect-containing supercells.
        
    \subsection{Formation energies and transition levels of defects and complexes}

	    Formation energies of the defects ($\Delta H_f^q$) as a function of the electron chemical potential $E_F$  in the band gap were calculated using the standard approach:
	    \begin{eqnarray}
    	    \Delta{H_{f}^{q}}(E_F)=&&E_{tot}^{q}-\sum_{i}n_{i}\mu_{i}+q(E_{VBM}+E_{F})+\nonumber\\&&+\Delta{E_{corr}},
	    \end{eqnarray}
	    where the Fermi level $E_F$ is referenced to the valence-band maximum (VBM) of non--defective diamond, $q$ stands for the charge state of a defect, $E_{tot}^q$ is the total energy of defective supercell, and $\mu_i$ is the chemical potential of the corresponding atoms ($i=$ C, Si, N). The $\mu_C$ value is calculated from the total energy of a perfect diamond lattice, while $\mu_N$ is deduced from the gas phase N$_2$ and $\mu_{Si}$ from bulk silicon in the diamond structure. The last term $\Delta{E_{corr}}$ is the finite-size correction based on the Freysoldt correction scheme  \cite{Frey_1,Frey_2}. We found that the application of the Freysoldt correction procedure leads to nearly the same results as the commonly used Makov-Payne scheme \cite{Makov_PhysRevB.51.4014,Lany_PhysRevB.78.235104}. The thermodynamic transition levels $E^{q_{1}|q_{2}}$ for a selected point defect can be calculated as  \cite{Frey_3}:
	    \begin{eqnarray}
        	E^{q_{1}|q_{2}}=\frac{\Delta{H_{f}^{q_{1}}(E_{F}=0)}-\Delta{H_{f}^{q_{2}}(E_{F}=0)}}{q_{2}-q_{1}}.
	    \end{eqnarray}
	
	    To investigate complex formation, we calculated the binding energy of a nitrogen atom. We define the binding energy of nitrogen in SiN, SiNV, or SiN$_2$V centers referenced to the energy of a substitutional N atom, N$_{\rm C}$:
	    \begin{eqnarray}
    	    E_{b}=\Delta{H_{f}^{q1}({\rm Si}{\rm N}_{x}{\rm V}_{y})}+\Delta{H_{f}^{q2}({\rm N}_{{\rm C}})}-\nonumber\\-\Delta{H_{f}^{q3}({\rm Si}{\rm N}_{x+1}{\rm V}_{y})}+ \sum_{j}q_{j}E_{F}.
    \label{eq:Eb}
	    \end{eqnarray}
	    With this definition, a positive value of the binding energy indicates a tendency toward complex formation.
	
	    Regarding formation of the complexes, we assumed that mobile interstitial nitrogen (N$_i$) can be trapped by Si$_{\rm C}$, SiV, or SiNV and form SiN, SiNV, or SiN$_2$V centers, respectively.  We determined migration barriers to shed some light on the possible formation pathways of SiN, SiNV, and SiN$_2$V centers.
        A climbing-image nudged-elastic-band method (CI-NEB)\cite{Henkelman} was used to find the minimum energy pathway (MEP) for the interstitial nitrogen and single vacancy migration in diamond.
	
	    To calculate the potential energy surface (PES) of the excited states and the corresponding ZPL values we applied the constrained density functional theory ($\Delta$SCF) method \cite{Gali_PRL2009theory} implemented in the VASP code. This method allows determining the atomic relaxation energy upon optical excitation.
	
    \subsection{Hyperfine interactions}

	    We calculated the hyperfine tensor including the core spin polarization effect in the Fermi-contact term within the frozen-valence approximation, using the implementation of Sz{\'{a}}sz \emph{et al.} \cite{Szasz_PhysRevB.88.075202} In this case, the plane-wave cutoff energy was raised to 800 eV, while the atomic positions taken from structural relaxation with cutoff energy of 520~eV were kept fixed.
	
    \subsection{Vibrational properties}

	    We used density functional perturbation theory (DFPT) to calculate the phonon spectrum of the systems. Here we used the Perdew-Burke-Ernzerhof (PBE) \cite{PBE_PhysRevLett.77.3865, PBE_PhysRevLett.78.1396} GGA functional. It has been demonstrated that PBE is capable of accurately reproducing the experimental lattice constant and phonon spectrum of diamond, as well as their dependence on pressure and temperature \cite{Ivanova_PSS2013}. In fact, the calculated Raman mode of 1336 cm$^{-1}$ for the perfect 512-atom supercell is very close to the experimental value \cite{Nemanich_PRB1979} of 1332 cm$^{-1}$. We note that the application of the HSE06 hybrid functional would be extremely computationally demanding, since we allow all atoms in the supercell to vibrate; as a result, for lower symmetry defects there are more than 800 vibrational degrees of freedom.
        To calculate the quasi-local vibrational modes with high precision we applied a very strict $10^{-4}$ eV/{\AA} force convergence criterion.

        The normal modes were analyzed using the inverse participation ratio approach \cite{Zhang_PRB2011}.The magnitude of the IPR is defined as
	    \begin{eqnarray}
        	IPR = \sum_{i=1}^{N}({x}_{i}^4+{y}_{i}^4+{z}_{i}^4)\left[\sum_{j=1}^{N}({x}_{j}^2+{y}_{j}^2+{z}_{j}^2)\right]^{-2},
	    \end{eqnarray}
    	where $\textsl{N}$ denotes the number of atoms in the system, and $\textbf{u}_{i}=(x_i,y_i,z_i)$ is the normalized 3$\textsl{N}$-dimensional vector of the atomic displacements of the corresponding vibrational eigenmode.

    \subsection{Vibronic coupling}

        We calculated electron--phonon coupling matrix elements and investigated the PL spectrum within the theory of Huang and Rhys\cite{HR}. A complete computational implementation of HR theory was provided by Alkauskas {\it et al.} and successfully demonstrated for NV center in diamond\cite{Alkauskas_2014}. We adopted the same implementation to investigate the electron--phonon coupling and PL spectrum of SiNV$^0$.

	   \begin{figure*}[th]
        \includegraphics[width = 0.75\linewidth]{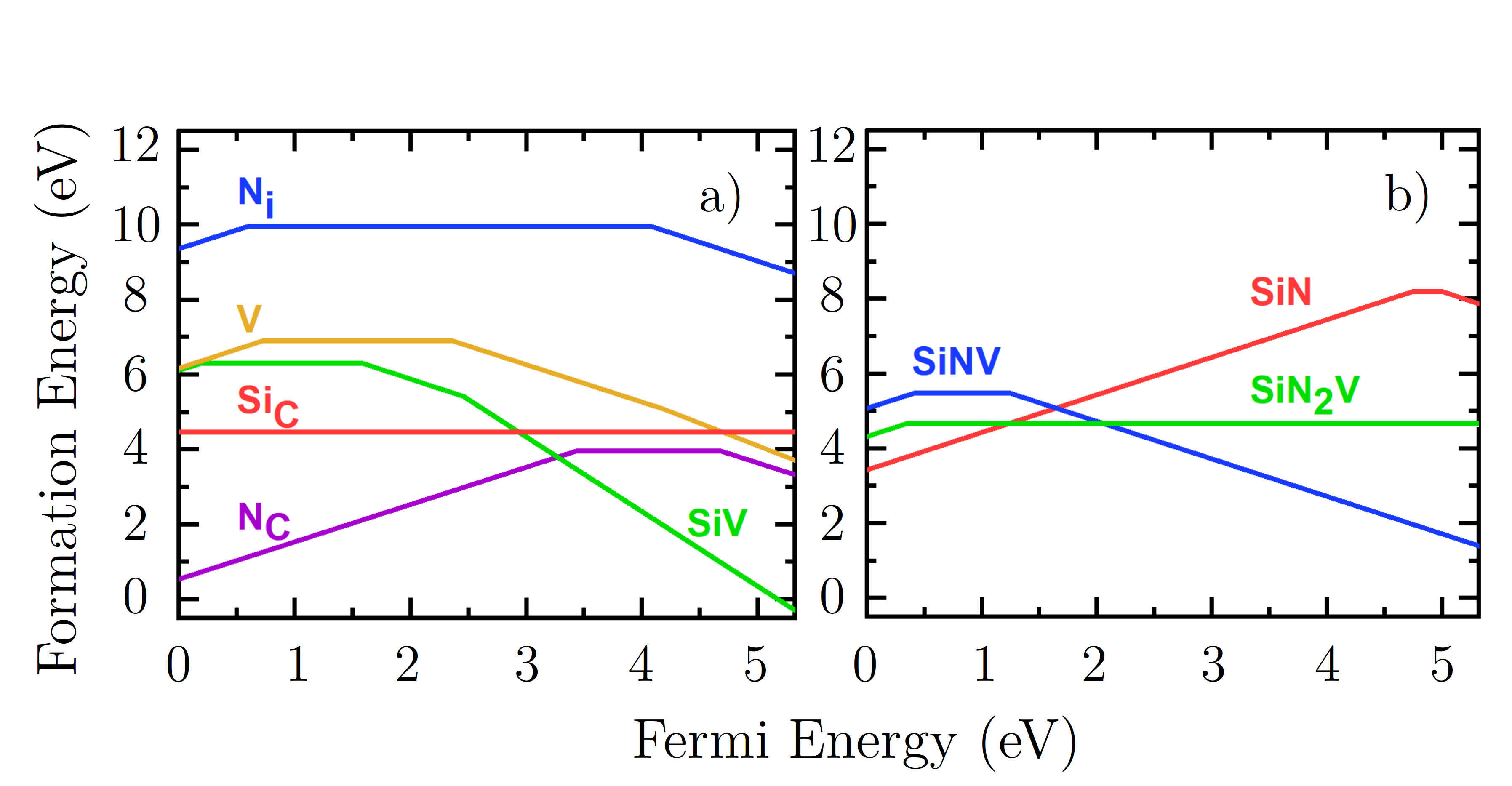}
        \caption{Formation energy of (a) previously investigated centers in diamond and (b) the complexes investigated in the present work, plotted as a function of Fermi level position in the fundamental band gap of diamond. The Fermi level is referenced to the VBM.
        \label{fig:Fig1}}
        \end{figure*}

As the determination of the absolute luminescence intensity is very complex, in this work the normalized luminescence intensity will be considered, which is defined as
	    \begin{eqnarray}
        L(\hbar\omega)=C\omega^{3}A(\hbar\omega),
	    \end{eqnarray}
	    with the optical spectral function $A(\hbar{\omega})$ defined as:
	    \begin{eqnarray}
	    \label{eq:A}
        A(\hbar\omega)=\sum_{m}\left|\braket{\chi_{gm}|\chi_{e0}}\right|^{2}\delta \left(E_{ZPL}-E_{gm}-\hbar\omega\right).
	    \end{eqnarray}
The $\chi_{e0}$ and $\chi_{gm}$ are vibrational levels of the excited and the ground state, and $E_{gm}$ is the energy of the $\chi_{gm}$ state, which is the sum of individual energies over all vibrational modes $k$: $E_{gm}=\sum_{k}n_{k}{\hbar\omega_{k}}$. A prefactor $\omega^{3}$ originates from the density of states of photons that cause the spontaneous emission $(\sim\omega^{2})$ and the perturbing electric field of those photons $(\vec{|\varepsilon|}^{2}\sim\omega)$, and $C$ is the normalization constant
	    \begin{eqnarray}
        C^{-1}=\int A(\hbar\omega)\omega^{3}d(\hbar\omega).
	    \end{eqnarray}
	    The most challenging part of Eq.~(\ref{eq:A}) is a set of multidimensional overlap integrals $\braket{\chi_{gm}|\chi_{e0}}$, that in practice, can be evaluated only for molecules, small clusters, or model defect systems. Vibrational modes that contribute to $A(\hbar\omega)$ originate from the bulk lattice vibrations as well as from localized or quasi-localized vibrational modes induced by defects. In addition, the normal modes in the excited state can, in principle, differ from those in the ground state \cite{Alkauskas_2014}. To simplify the evaluation of the spectral function $A(\hbar{\omega})$ (Eq.~(\ref{eq:A})), we made the assumption that the normal modes that contribute to the luminescence line shape are identical in the excited and ground state. Although this assumption does not strictly hold for the NV center, the agreement between numerical and experimental results is impressive \cite{Alkauskas_2014}.
In the case of SiNV$^0$, the assumption is expected to be better justified, since the symmetry of the ground and excited state remains unchanged and the calculated Stokes shift associated with the a' $\rightarrow$ a'' internal optical transition is two times smaller than the Stokes shift of the $a_{1}$ $\rightarrow$ $e$ transition in NV center.
	
	    Assuming that the phonon modes in the excited and ground state are identical, the optical spectral function defined in Eq.~(\ref{eq:A}) can be rewritten using a generating function method suggested by Lax\cite{Lax}, Kubo and Toyozawa\cite{Kubo}. A very important quantity that should be determined is the spectral function of the electron--phonon coupling\cite{Miyakawa}, which is expressed as
	    \begin{eqnarray}
	    \label{eq:B}
            S(\hbar\omega)=\sum_{k}S_{k}\delta \left(\hbar\omega-\hbar\omega_{k}\right),
	    \end{eqnarray}
	    where the sum goes over all phonon modes $k$ with frequencies $\omega_{k}$. $S_{k}$ is defined as the partial Huang-Rhys factor, given by\cite{MARKHAM}
	    \begin{eqnarray}
	    \label{eq:C}
            S_{k}=\frac{\omega_{k}q_{k}^{2}}{2\hbar} \, ,
	    \end{eqnarray}
	    where $q_{k}$ is defined as
	    \begin{eqnarray}
	    \label{eq:D}
            q_{k}=\sum_{\alpha,i}\sqrt{m_{\alpha}}\left(R_{\alpha,i}^{(e)}-R_{\alpha,i}^{(g)}\right)\Delta r_{k,\alpha,i} \, .
	    \end{eqnarray}
$m_{\alpha}$ is the mass of atom $\alpha$, $R_{\alpha,i}^{(e)}$ and $R_{\alpha,i}^{(g)}$ are equilibrium atomic coordinates of atom $\alpha$ along the direction $i$ in the excited state and the ground state, and $\Delta r_{k,\alpha,i}$ is the normalized displacement vector of atom $\alpha$ along the direction $i$ in the phonon mode $k$. Having obtained the function $S(\hbar\omega)$, one can express the optical spectral function in terms of the Fourier transform of the generating function $G(t)$:
	    \begin{eqnarray}
	    \label{eq:E}
            A(E_{ZPL}-\hbar\omega)=\frac{1}{2\uppi}\int_{-\infty}^{\infty}G(t)e^{i\omega t-\gamma |t|}dt,
	    \end{eqnarray}
	    where $\gamma$ represents the width of the ZPL. The generating function is defined as
	    \begin{eqnarray}
	    \label{eq:F}
            G(t)= e^{S(t)-S(0)},
	    \end{eqnarray}
	    where $S(t)$ is the Fourier transform of the partial HR factor
	    \begin{eqnarray}
	    \label{eq:G}
            S(t)=\int_{0}^{\infty}S(\hbar\omega)e^{-i\omega t}d(\hbar\omega).
	    \end{eqnarray}
	    $S(0)$, in turn, is the total HR factor for a given optical transition and is defined as the sum of the partial HR factors over all phonon modes $k$:
	    \begin{eqnarray}
	    \label{eq:H}
            S(0)=\int_{0}^{\infty}S(\hbar\omega)d(\hbar\omega)=\sum_{k}S_{k}.
	    \end{eqnarray}
	
	    The total HR factor is an important parameter that describes the vibrational structure of the PL band. It is commonly used to determine the weight of the ZPL\cite{Davies,MARKHAM,Lax}
	    \begin{eqnarray}
	    \label{eq:I}
            w_{ZPL}=e^{-S(0)}.
	    \end{eqnarray}
	    The quantity $w_{ZPL}$ is often called the Debye-Waller (DW) factor and it describes the fraction of photons emitted into the ZPL.
	
	    \section{RESULTS AND DISCUSSION}

	    \subsection{Formation of Si--N--V defects and their stability in diamond}

        \begin{figure*}[t]
        \includegraphics[width = 1.5\columnwidth]{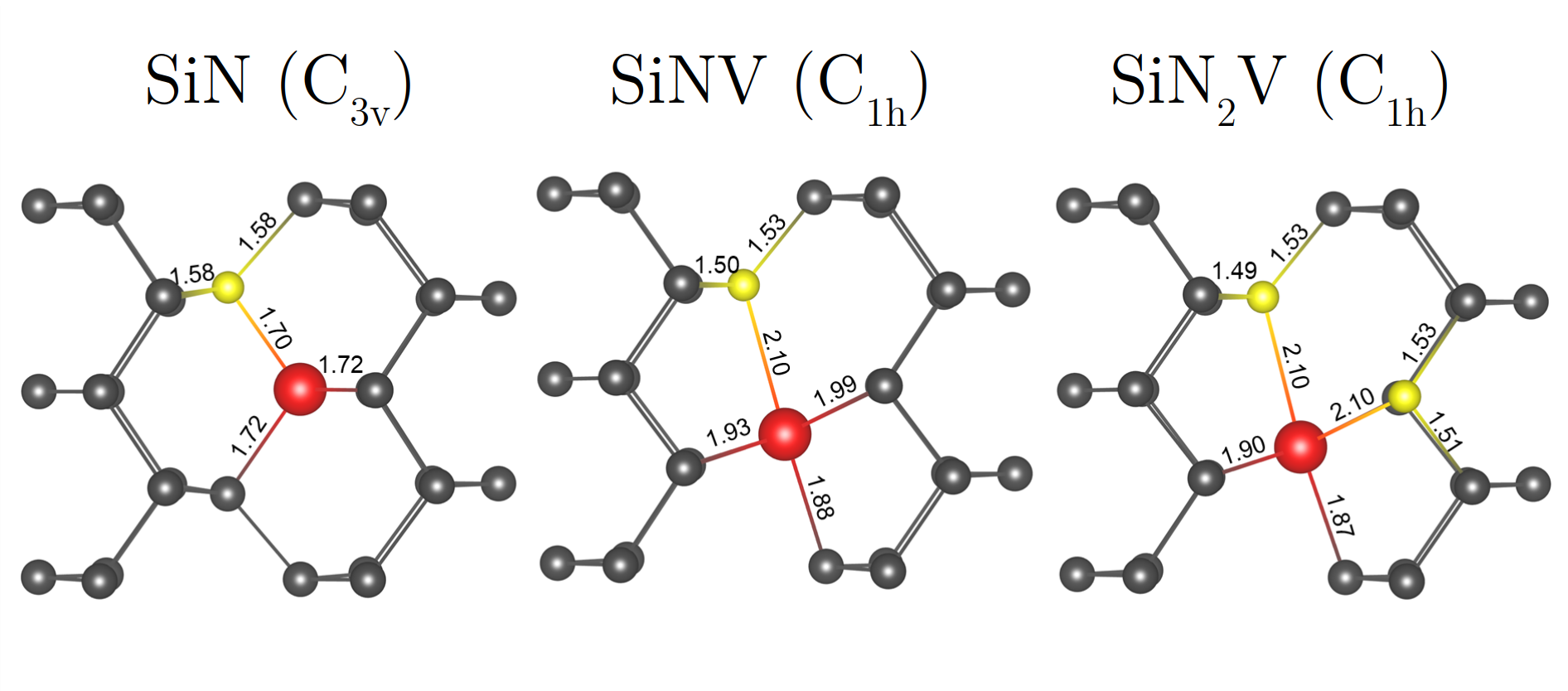}
        \caption{Ground-state geometries of the investigated complexes in the neutral charge states extracted from a 512--atomic supercell together with the point-group symmetry labels. The Si atom is represented by a large red sphere and the nitrogen atoms by small yellow spheres. Bond lengths are given in angstrom (\AA).}
        \label{fig:Fig2}
        \end{figure*}

        To shed light on the processes associated with the formation of SiN, SiNV, and SiN$_2$V centers in diamond, we calculated their formation energies as a function of Fermi level (referenced to the VBM).
        We also calculated the formation energies of various intermediate defects. The results are shown in Fig.~\ref{fig:Fig1}.
        Formation energies of the defects shown in Fig.~\ref{fig:Fig1}(a) have been calculated by other groups, and our results agree with those reports as well as with experimental data\cite{Jones_DRM2015,Newton_DRM2002,Deak2014, Evans}.

         The formation energies of the SiN, SiNV, and SiN$_2$V centers are shown in Fig.~\ref{fig:Fig1}(b). Based on these diagrams, one can read off the relative stability of a given complex in various charge states with respect to other defects or complexes. Figure~\ref{fig:Fig1} shows that the investigated complexes are electrically active point defects and can in principle exist in positive, neutral, and negative charge states.

        The SiN dimer embedded in the diamond lattice is stable in a positive charge state over most of the range of Fermi levels, and exhibits a $E^{+|0}$ level at 0.57 eV below the CBM and a $E^{0|-}$ level at 0.32 eV below the CBM.  The electrical activity of this center will be discussed in the next section (Sec.~\ref{sec:SiN}).
        The SiNV defect has an acceptor level at 1.24~eV and is mostly stable in the negative charge state. The SiN$_2$V center, finally, has a narrow stability window near the  VBM corresponding to SiN$_2$V$^{+}$, but is predominantly stable as a neutral defect.
        
        The relaxed structures of these complexes, together with the point-group symmetry labels, are shown in Fig.~\ref{fig:Fig2}.
        We note that, while the geometries are shown for the neutral charge state, the symmetry labels apply to all charge states of the respective complexes.
        In general, the relaxations of the nearest neighbor atoms around the defects are smaller than 0.1 {\AA}. The largest relaxation is observed for the SiN dimer: the nitrogen atom moves outward by $\sim$0.17 {\AA} along  the [111] direction. As can be seen in Fig.~\ref{fig:Fig2}, the Si atom belonging to SiNV and SiN$_2$V centers no longer sits midway between two C atoms (the split-vacancy position that occurs in the case of the SiV center) but moves slightly away from the nitrogen neighbors.

        \begin{figure}[t]
        \includegraphics[width = 1.0\columnwidth]{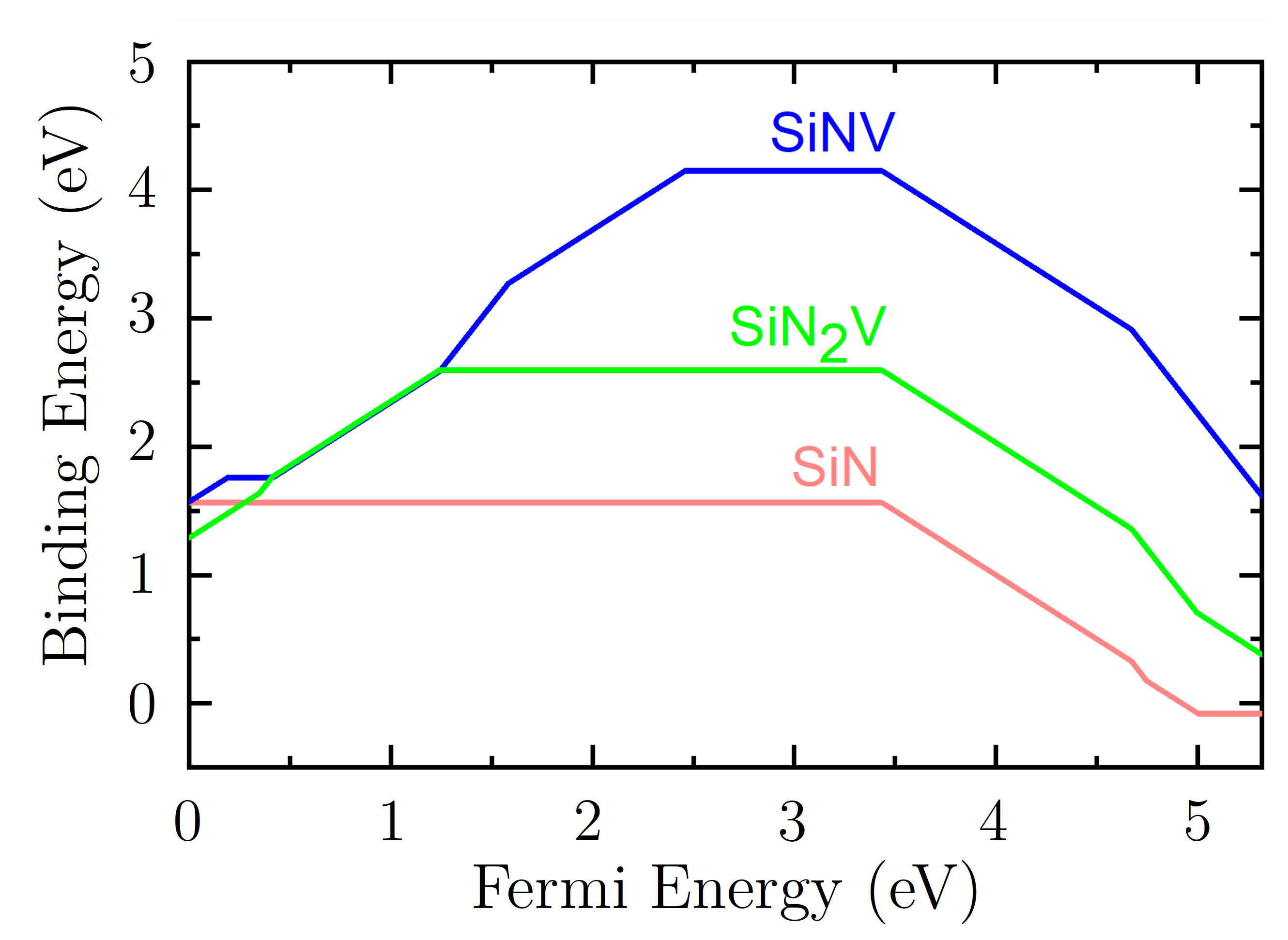}
        \caption{Calculated binding energy of the nitrogen atom to Si, SiV, and SiNV complexes as a function of Fermi-level position in the fundamental band gap of diamond. The Fermi level is referenced to the VBM.}
        \label{fig:Fig3}
        \end{figure}

    To evaluate the tendency toward complex formation, we calculated the binding energy of a nitrogen atom to substitutional Si, to SiV, and to SiNV complexes, using the definition of Eq.~(\ref{eq:Eb}).  The results are shown in Fig.~\ref{fig:Fig3}; positive values of the calculated binding energies indicate a thermodynamic driving force for complex formation.
The binding energies are largest when the charge of the complex equals the total charge of the constituents (horizontal line segments in Fig.~\ref{fig:Fig3}).   This is also the physically plausible situation, since it is unlikely that charge can be exchanged during the complex formation.  The relevant Fermi-level positions include values near the center of the gap, where the Fermi level is likely to be placed due to the damage induced by irradiation or implantation.
We see that in all three cases it is more favorable to place a nitrogen atom in the complex relative to having nitrogen present as an isolated N$_{\rm C}$.
    These binding energies are also large enough to suppress dissociation of the complexes, once formed.  We note that the binding energy is especially large in the case of the SiNV complex, which will be of particular interest.
    In the case of defects containing a single vacancy, the binding energy is roughly twice as high as the binding energy of the SiN defect. The stronger driving force for nitrogen complexing with SiV and SiNV originates partially from the electrostatic attraction between the positively charged substitutional nitrogen and the negatively charged vacancy-containing complex.

        \begin{figure*}[t]
        \centering
        \includegraphics[width = 1.5\columnwidth]{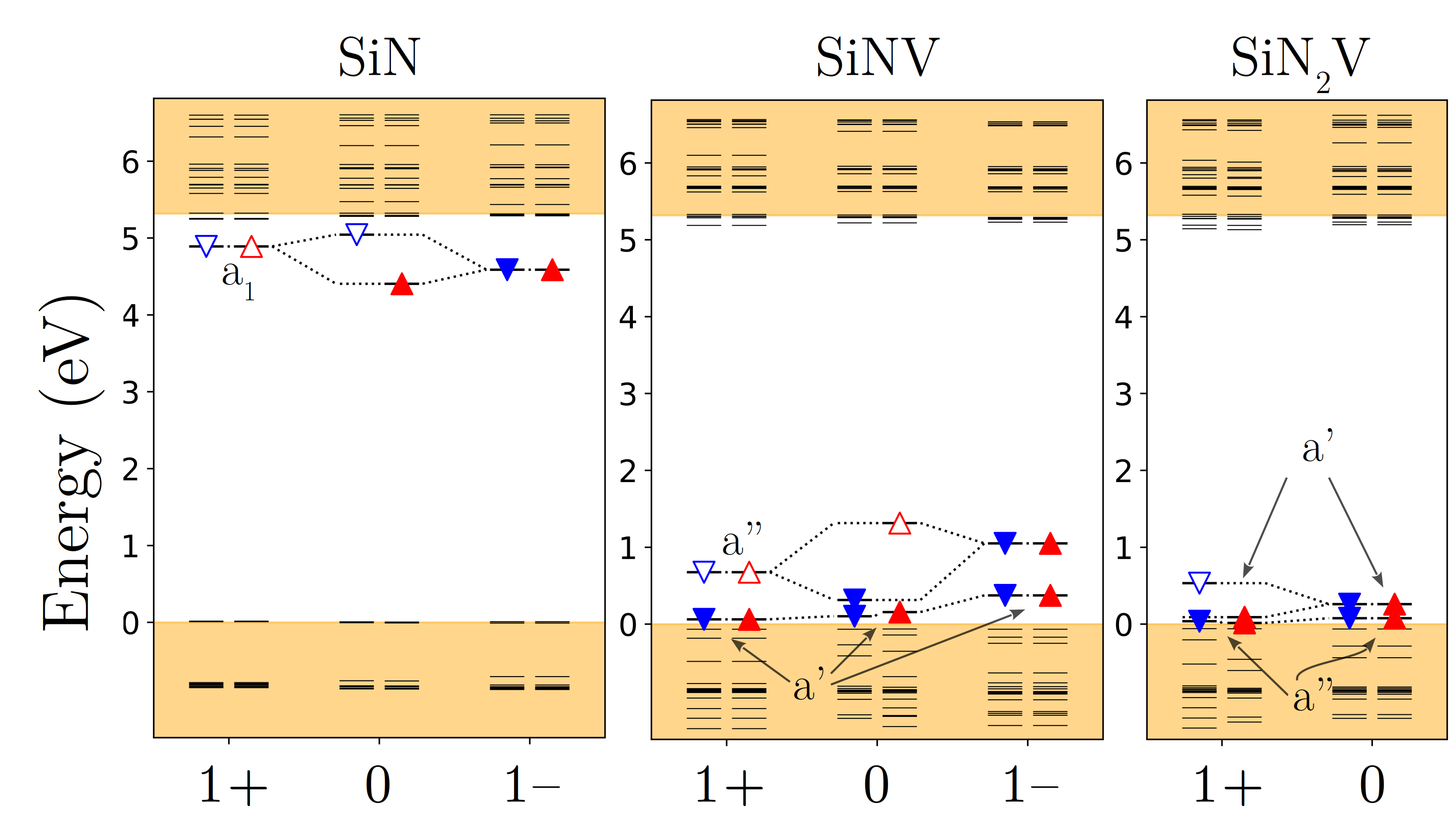}
        \caption{Localized Kohn--Sham levels of the investigated SiN, SiNV, and SiN$_2$V complexes in diamond calculated with the HSE06 hybrid functional. The light orange shaded area represents the conduction and valence bands of bulk diamond. Spin-down (-up) channels are indicated by blue (red) triangles, and filled (unfilled) triangles represent the occupied (empty) states. For the sake of clarity, the corresponding KS states in different charge states are linked by black dotted lines. All the localized defect-related states are labeled with their symmetry representation.}
        \label{fig:Fig4}
        \end{figure*}

        We also briefly analyzed the possible pathways leading to the formation of these centers in diamond.
        Experimental studies on nitrogen aggregation and formation mechanisms of N-related defects in diamond have revealed two mobile species: the neutral vacancy and interstitial nitrogen. Our calculated formation energies $\Delta H_{\rm V}^{0}=6.9$ eV, $\Delta H_{\rm N_{i}}^{0}=9.9$ eV, and migration barriers $\Delta E_{\rm V}^{0}=2.7$ eV, $\Delta E_{\rm N_{i}}^{0}=1.7$ eV for these species, are close to previously calculated values $\Delta H_{\rm V}^{0}=7.14$ eV\cite{Deak2014}, $\Delta H_{\rm N_{i}}^{0}=9.4\div10.4$ eV\cite{Jones_DRM2015} and $\Delta E_{\rm V}^{0}=2.3\pm0.3$ eV\cite{Deak2014}, $\Delta E_{\rm N_{i}}^{0}=1.8$ eV\cite{Jones_DRM2015}.

        \begin{figure*}[t]
        \includegraphics[width = 1.5\columnwidth]{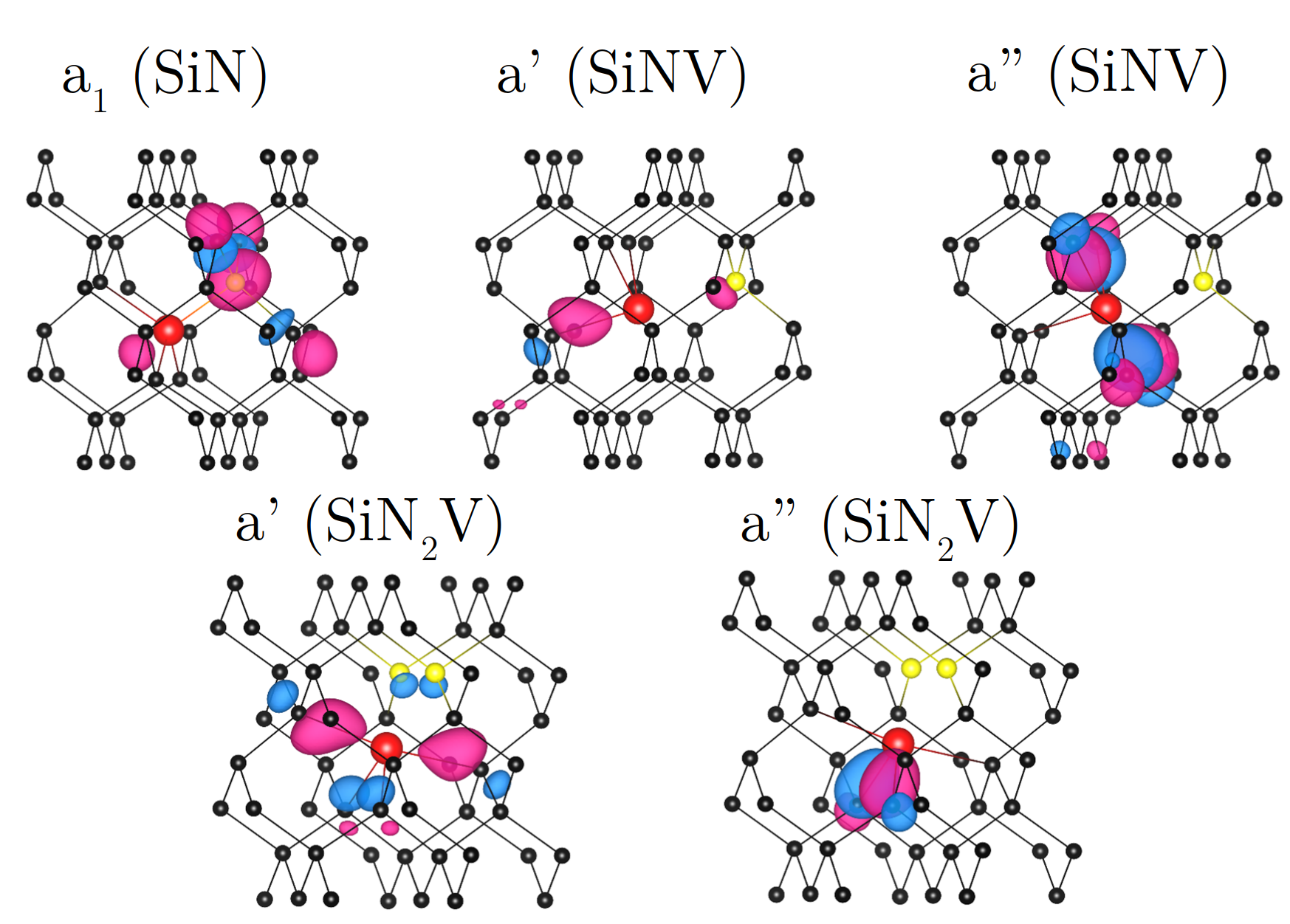}
        \caption{3D representation of the calculated Kohn--Sham wave functions for Si--N--V complexes in diamond. The red (blue) lobes indicate the positive (negative) phase of the wave functions with an arbitrarily selected isosurface value. The silicon, nitrogen, and carbon atoms are represented by red, yellow, and black spheres.}
        \label{fig:Fig5}
        \end{figure*}

        As shown in Fig.~\ref{fig:Fig3}(a), the formation energies of interstitial nitrogen and vacancies are very high, and they are thus unlikely to form under equilibrium conditions.  We thus assume they are generated in nonequilibrium fashion, \emph{via} ion implantation or irradiation, and subsequently subjected to annealing.
        The formation of SiN, SiNV, and SiN$_2$V centers then involves self-diffusion of vacancies and diffusion of nitrogen \cite{Jones_DRM2015}. The vacancy migration energy of 2.7 eV is low enough to initiate diffusion at about 800\degree C, and vacancies can be trapped by substitutional N and Si.
        In the case of interstitial nitrogen diffusion the energy barrier of 1.7~eV is even smaller.
        After heating up the sample above 600\degree C, the diffusing interstitial N will eventually be trapped by V, N$_{\rm C}$ or SiV, forming N$_{\rm C}$, N$_{2}$ (a pair of N atoms sharing a lattice site), or SiN.\cite{Jones_DRM2015}
     Alternatively, nitrogen-containing complexes could be formed directly during irradiation, by introduction of nitrogen directly next to a Si atom or SiV complex, as was previously suggested for formation of NV centers in Ref.~\onlinecite{Deak2014}.  Subsequent temperature treatment would still be necessary to anneal out damage.

         Interestingly, the neutral SiNV center was recently detected experimentally for the first time by Breeze {\it et al}\cite{breeze}. Annealing the samples for 1h at different temperatures and using EPR and IR spectroscopy the increased tendency towards formation of SiNV$^{0}$ has been revealed.
        Therefore, the simultaneous presence of N and Si atoms introduced to the diamond lattice in nonequilibrium fashion followed by proper heat treatment enables the formation of Si--N--V complexes.

        \subsection{The SiN center as an active donor in diamond}
        \label{sec:SiN}

    The SiN dimer embedded in the diamond lattice exhibits a SiN$^{+}$ $\rightarrow$ SiN$^{0}$ transition corresponding to donor level at 0.57~eV below the CBM. This value is almost the same as the activation energy of the phosphorus donor, which is commonly regarded as the shallowest individual donor dopant in diamond \cite{Kato_DRM2005, Czelej_PhysRevB.98.075208}.
        In contrast, substitutional nitrogen is known as a deep donor with activation energy $\sim$1.9~eV \cite{Weber2010, Czelej_Diamond}.
        The silicon atom itself is isoelectronic and isostructural with carbon, and hence, electrically inactive, but the presence of Si in the SiN dimer pushes the donor level significantly towards the CBM. As a consequence, silicon may turn inactive nitrogen into a shallower donor and diamond co-doped with Si and N may exhibit $n$-type conductivity.

        We note that, in addition to the $E^{+|0}$ level, the SiN center also has an  $E^{0|-}$ level at 0.32 eV below the CBM.  The presence of this level should not interfere with inducing $n$-type conductivity, since at the temperatures where the center is ionized (\emph{i.e.}, where the ionization energy of 0.57 eV is overcome), the occupation of this negative charge state will be negligible.

        In light of ongoing research efforts to achieve $n$-type diamond, our result could be very important; a systematic experimental investigation is called for.

        \subsection{Electronic structure of Si--N--V defects in diamond}

        We now examine the quantum behavior as well as optical and magnetic features of the investigated defects by calculating the Kohn--Sham eigenvalue spectrum and analyzing defect-induced gap states using group theory (see Fig.~\ref{fig:Fig4}). To facilitate experimental identification, we calculated the EPR signals of hyperfine coupling for all the investigated paramagnetic centers.   Hyperfine structure calculations were carried out for the most abundant and stable magnetic isotopes of carbon $^{13}$C ($I=1/2$; abundance = 1.1\%), silicon $^{29}$Si ($I=1/2$; abundance = 4.7\%), and nitrogen $^{14}$N ($I=1$; abundance = 99.6\%). The results are listed in Table~\ref{tab:hyperfine00}).

        To facilitate further experimental identification, particularly of singlet states that cannot be captured by EPR, we calculated the vibrational spectra, which can be detected in infrared absorption or in the phonon sideband in a PL experiment.  Values are given in Table~\ref{tab:modes}.
        Quasi-local vibrational modes were identified using the IPR analysis described in Sec.~\ref{sec:theory}.

        \begin{table}[t]
        	\caption{\label{tab:hyperfine00}
        		Hyperfine constants of the Si--N--V complexes calculated with the HSE06 functional.
                $R$ (MHz) is the magnitude of the hyperfine constants, and $\theta$ (\textdegree), $\varphi$ (\textdegree) provide the direction in spherical coordinates. The $\theta$ value describes the polar angle referenced to the $[001]$ direction and $\varphi$ is the azimuthal angle referenced to the $[100]$ direction [in the $(001)$ plane].}
            	\begin{ruledtabular}
		        \begin{tabular}{ lrrrrrrrrrr }
        			&              &\multicolumn{3}{c}{$A_{xx}$}&\multicolumn{3}{c}{$A_{yy}$}&\multicolumn{3}{c}{$A_{zz}$}\\
        			&            &$R$&$\theta$&$\varphi$&$R$&$\theta$&$\varphi$&$R$&$\theta$&$\varphi$\\	\hline 			
        		\\[-2.5mm]
        		\parbox[t]{3mm}{\multirow{2}{*}{\rotatebox[origin=c]{90}{SiN$^0$}}}
        		&{$^{29}$Si}&3&114\textdegree&333\textdegree&8&55\textdegree&45\textdegree&3&58\textdegree&289\textdegree\\
        		&{$^{14}$N}&367&66\textdegree&153\textdegree&374&55\textdegree&45\textdegree&367&66\textdegree&296\textdegree\\
        		\hline
        		\\[-2.5mm]
        		\parbox[t]{3mm}{\multirow{6}{*}{\rotatebox[origin=c]{90}{SiNV$^{0}$}}}
        		&{1 $^{13}$C}&94&122\textdegree&134\textdegree&43&119\textdegree&24\textdegree&43&46\textdegree&81\textdegree\\
        		&{2 $^{13}$C}&94&122\textdegree&316\textdegree&43&134\textdegree&189\textdegree&43&61\textdegree&246\textdegree\\
                &{3 $^{13}$C}&124&122\textdegree&316\textdegree&55&140\textdegree&177\textdegree&56&69\textdegree&240\textdegree\\
        		&{4 $^{13}$C}&124&122\textdegree&135\textdegree&56&111\textdegree&31\textdegree&55&40\textdegree&93\textdegree\\
        		&{$^{29}$Si}&112&83\textdegree&45\textdegree&122&90\textdegree&315\textdegree&118&87\textdegree&225\textdegree\\
        		&{$^{14}$N}&5&62\textdegree&45\textdegree&4&90\textdegree&315\textdegree&4&152\textdegree&45\textdegree\\
                \hline
        		\\[-2.5mm]
                \parbox[t]{3mm}{\multirow{4}{*}{\rotatebox[origin=c]{90}{SiN$_2$V$^{+}$}}}
        		&{$^{13}$C}&33&63\textdegree&210\textdegree&96&123\textdegree&139\textdegree&32&45\textdegree&90\textdegree\\
        		&{$^{29}$Si}&47&82\textdegree&189\textdegree&63&134\textdegree&107\textdegree&64&45\textdegree&90\textdegree\\
        		&{1 $^{14}$N}&8&55\textdegree&29\textdegree&4&104\textdegree&309\textdegree&4&38\textdegree&237\textdegree\\
        		&{2 $^{14}$N}&8&67\textdegree&142\textdegree&4&41\textdegree&22\textdegree&4&59\textdegree&247\textdegree\\
	    	    \end{tabular}
	        \end{ruledtabular}
        \end{table}

  \begin{table}[b]
         \caption{\label{tab:modes} Calculated frequencies and inverse participation ratios (IPR) for the quasi-local vibration modes of the Si--N--V complexes.}
            \centering
            \begin{ruledtabular}
            \begin{tabular}{lccc}
            Defect & Symmetry label & Frequency (cm$^{-1}$) & IPR \\
            \hline
            \multirow{2}{*}{SiN} & $e$ & 477 & 0.064 \\ & $a_1$ & 450 & 0.047 \\
            \hline
            \multirow{2}{*}{SiNV} & $a$' & 480 & 0.116 \\ & $a$'' & 435 & 0.055 \\
            \hline
            \multirow{2}{*}{SiN$_2$V} & $a$' & 445 & 0.044 \\ & $a$'' & 484 & 0.055 \\ & $a$'' & 506 & 0.014 \\
            \end{tabular}
            \end{ruledtabular}
        \end{table}

 \subsubsection{The SiN center}
 \label{sec;elstr_SiN}

The ground state of the SiN center in diamond exhibits C$_{3v}$ symmetry. Under the C$_{3v}$ crystal field, there are two possible combinations of molecular orbitals: totally symmetric nondegenerate a$_1$ or a$_2$, and doubly degenerate e. Only the a$_1$ state appears in the band gap, close to the CBM, whereas the e state falls deep within the valence band and loses its local character. As can be seen in Fig.~\ref{fig:Fig5}, the a$_1$ orbital is predominantly localized on the nitrogen atom. In the positive charge state, the a$_1$ orbital is empty and the system has a singlet $S$~=~0 spin state. in the neutral charge state, an additional electron occupies the a$_1$ orbital forming a doublet $S$~= $1/2$ spin state that can in principle be detected with EPR spectroscopy.
Table~\ref{tab:hyperfine00} shows that in the case of SiN$^0$, significant hyperfine interaction with the $^{14}$N nuclear spin arises from strong spin density localization on the a$_1$ orbital.
It is intereesting to compare the hyperfine constants of SiN$^0$ with those of the isolated N$_{\rm C}$ impurity in diamond.
For  N$_{\rm C}^0$ (the so-called P1 center), the experimental values are $A_{\parallel}$ = 146 MHz and $A_{_\bot}$ = 173 MHz \cite{Kamp2018}. This is approximately half of the values for the SiN$^0$ defect (Table~\ref{tab:hyperfine00}).  The difference indicates stronger localization of spin density in the vicinity of the nitrogen atom in the case of the SiN center. This may be explained by the electronegativity difference between N--Si (1.14 on the Pauling scale\cite{Pauling}) and N--C (0.49 on Pauling scale\cite{Pauling}): nitrogen in the SiN center polarizes the charge density much more strongly than in the P1 center, which in turn leads to the stronger localization of spin density in the vicinity of the N atom and stronger hyperfine coupling.

        When an excess electron is introduced to the system, the center becomes negatively charged and the a$_1$ becomes entirely occupied, resulting in a closed-shell spin singlet configuration. The negatively charged SiN$^{-}$, however, has a narrow stability window close to the CBM and under typical conditions it is unlikely to occur.

        We also calculated the vibrational spectrum of the SiN center. As shown in Table~\ref{tab:modes},
        the quasi-local vibrational modes associated with the SiN dimer occur at 477 cm$^{-1}$ (doubly degenerate e mode with the IPR = 0.064) and 450 cm$^{-1}$ (nondegenerate a$_1$ mode with the IPR = 0.047).

        As the SiN defect in diamond induces only one nondegenerate a$_1$ state in the band gap, there are no intra-defect optical transitions capable of producing photoluminescence. Nevertheless, the accessible a$_1$ $\rightarrow$ CBM transition enables the photo-ionization of SiN with near-infrared light.

 \subsubsection{The SiNV center}

        We now discuss the SiNV center, which can be generated, for instance, by trapping a mobile vacancy.
         The SiNV complex has C$_{1h}$ symmetry (see Fig.~\ref{fig:Fig2}).
         Two defect-related localized states appear in the band gap: the symmetric a' and the asymmetric a'' (Fig.~\ref{fig:Fig4}).
         The 3D plot of a' and a'' orbitals (Fig.~\ref{fig:Fig5}) indicates that the former is localized on a carbon dangling bond and partially on nitrogen, whereas the latter is a pure C dangling-bond state.

         In the positive charge state the a' state is fully occupied and the a'' state is empty, leading to a singlet $S$~=~0 ground state.
         The neutral SiNV$^0$ is a paramagnetic doublet S~=~$1/2$ with a half-filled a'' orbital (see Fig.~\ref{fig:Fig4}).
         The spin density of SiNV$^0$ is predominantly localized on C dangling bonds, leading to substantial hyperfine coupling with $^{13}$C (see Table~\ref{tab:hyperfine00}). Trapping of an excess electron leads to the SiNV$^{-}$ charge states, which forms a closed-shell singlet ($S$~=~0) which is  stable over a wide range of Fermi energy levels (see Fig.~\ref{fig:Fig3}).
         We found two quasi-local vibrational modes for this complex (see Table~\ref{tab:modes}). The calculated frequencies and corresponding IPR values for the a' symmetric stretching and a'' bending modes are 480 cm$^{-1}$ (IPR=0.116) and 435 cm$^{-1}$ (IPR=0.055), respectively.

        \begin{figure}[t]
        \centering
        \includegraphics[width=0.75\columnwidth]{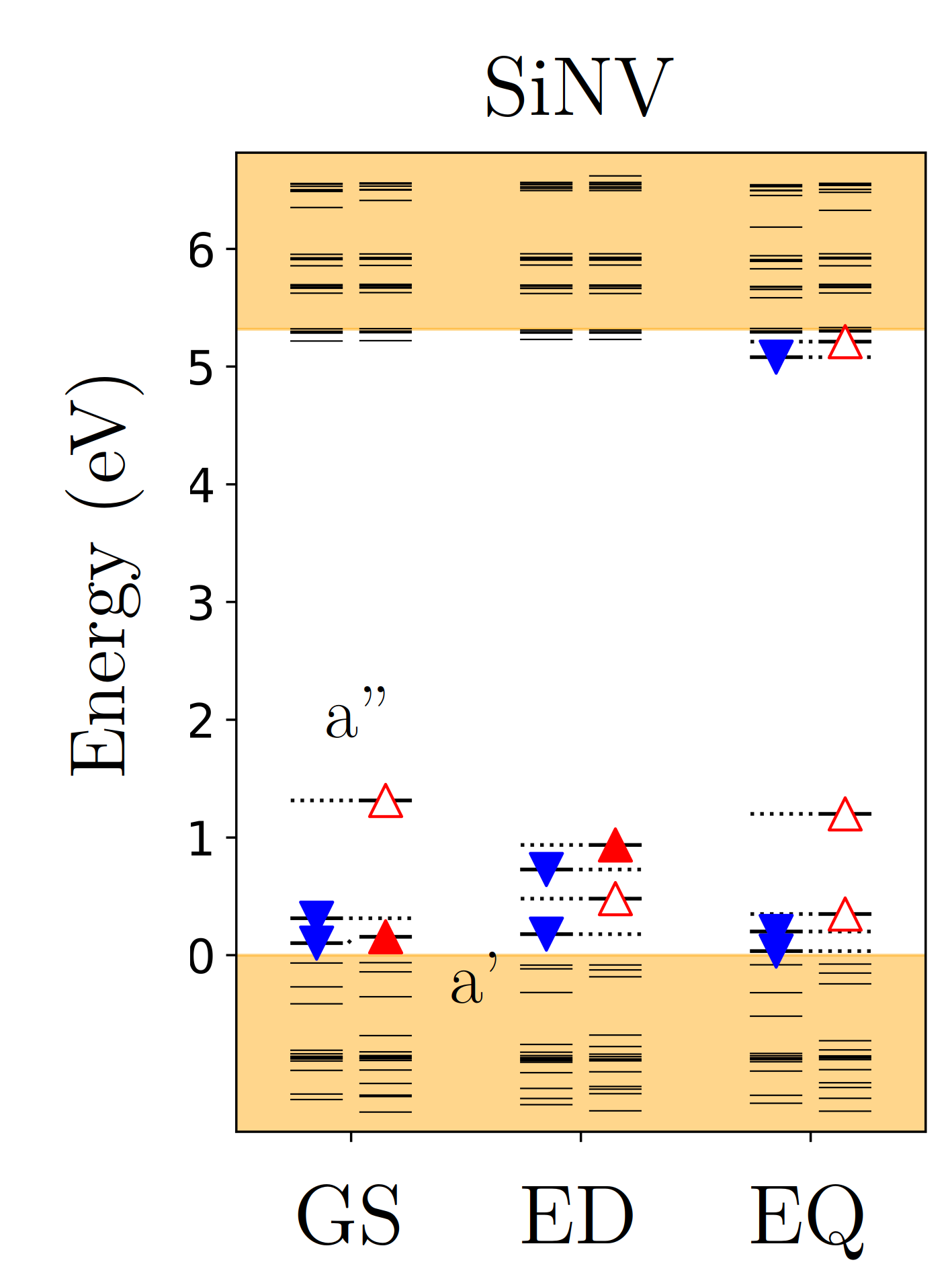}
        \caption{Localized Kohn--Sham levels of SiNV center in diamond calculated with the HSE06 hybrid functional. The ground state (GS), the excited state doublet (ED), and the excited state quartet (EQ) are presented. The  light orange shaded area represents the conduction and valence band of bulk diamond. Spin-down (-up) channels are indicated by blue (red) triangles, and filled (unfilled) triangles represent the occupied (empty) states.}
        \label{fig:Fig6}
        \end{figure}

        \begin{figure}[b]
        \centering
        \includegraphics[width=1.0\columnwidth]{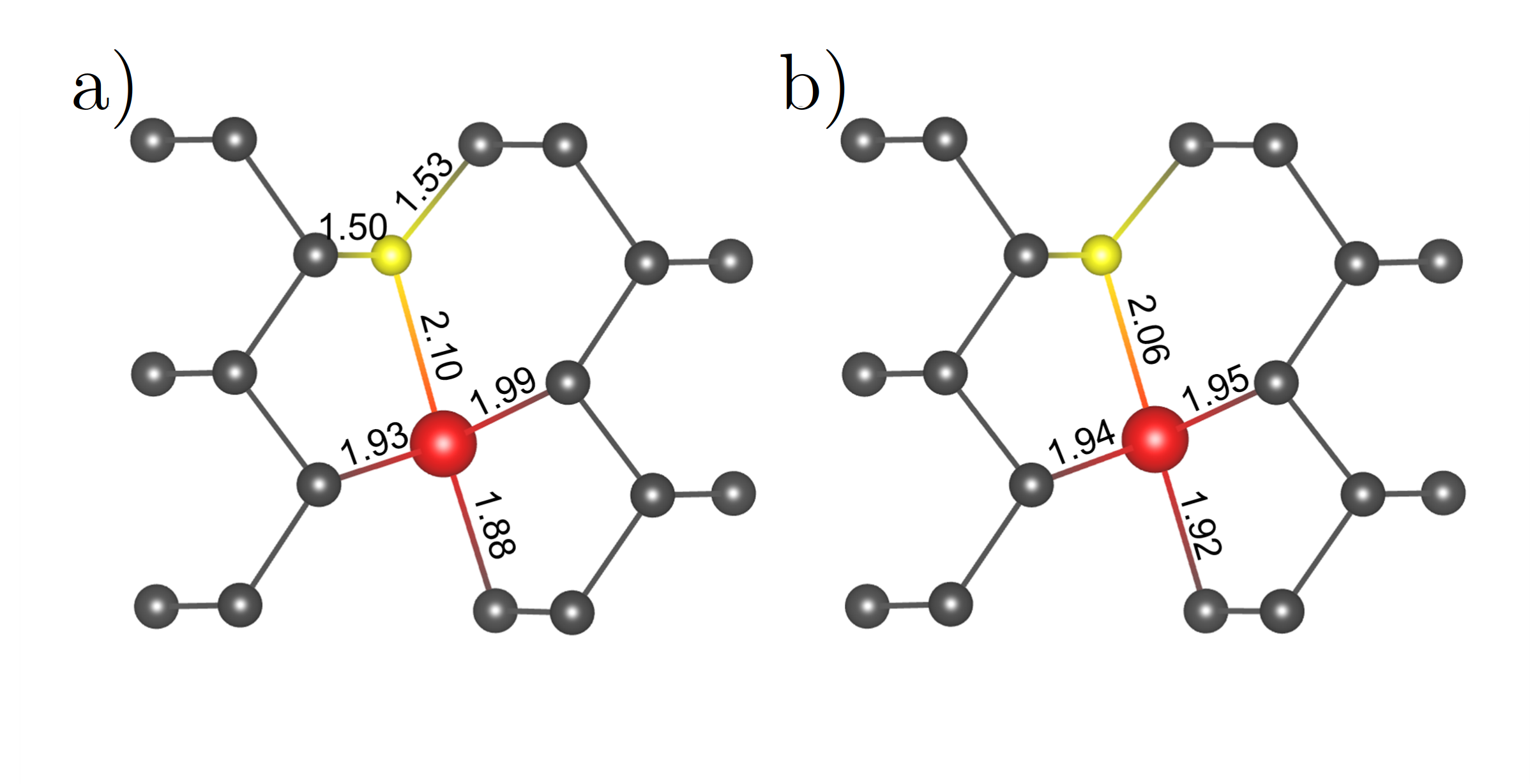}
        \caption{The relaxed structure of SiVN$^0$ center in a) the ground state; b) the lowest excited state. The Si atom is represented by a large red spheres and the nitrogen atoms by small yellow spheres. Bond lengths are given in angstrom (\AA).}
        \label{fig:Fig7}
        \end{figure}

          \subsubsection{The SiN$_2$V center}

        Finally, we analyzed the electronic structure of the SiN$_2$V center. Similar to SiNV, two defect-related a' and a'' states appear in the band gap.
        However, as seen in Fig.~\ref{fig:Fig4}, these states are placed very close to the valence band and may resonate with it.
        The two nitrogen atoms present in the defect structure donate two additional electrons, completely filling entirely the gap states in the neutral charge state.  SiN$_2$V$^0$ thus has a very stable closed-shell singlet $S$~=~0 configuration.

        The positively charged SiN$_2$V$^{+}$ is paramagnetic doublet (S~=~$1/2$) but its stability window is very narrow (see Fig.~\ref{fig:Fig3}). Nevertheless, we calculated hyperfine coupling constants for this doublet and as can be seen in Table~\ref{tab:hyperfine00} the coupling is weak, confirming that spin density is less localized in the vicinity of the defect.

        The prevailing neutral charge state, SiN$_2$V$^0$, is nonparamagnetic and, due to a lack of empty gap states, its optical excitation to the CBM or higher excited states would involve deep UV light. The center can potentially be experimentally detected by its infrared or Raman fingerprint. We identified three quasi-local vibrational modes associated with SiN$_2$V$^0$: a symmetric stretching a' mode at 445 cm$^{-1}$ (IPR=0.044), an asymmetric stretching mode at 484 cm$^{-1}$ (IPR=0.055), and a bending mode at 506 cm$^{-1}$ (IPR=0.014).

        \subsection{Electron--phonon coupling and quantum emission from SiNV$^{0}$}

        For the SiNV center, we focused our attention on the neutral charge state, because it is paramagnetic, and therefore, of interest for quantum information applications.
        In fact, SiNV$^0$ has been experimentally detected by EPR spectroscopy\cite{breeze}.
        We now discuss the fact that the SiNV defect exhibits an interesting internal a' $\rightarrow$ a'' optical transition in the neutral charge state, and we will demonstrate that SiNV$^0$ can be a robust single-photon source.

        First, we analyze the electronic structure of the excited-state configurations of SiNV$^0$ (see Fig.~\ref{fig:Fig6}).
       In the ground state, the neutral SiNV$^0$ center is a paramagnetic doublet S~=~$1/2$ with a half-filled a'' orbital.
        Spin--conserving coherent excitation of the defect promotes an electron from the a' to the a'' orbital in the spin-minority channel, leaving behind one hole in the a' orbital.
        Theoretically, by flipping over the electron spin one would generate a high-spin quartet ($S=3/2$) electronic configuration. Our $\Delta$SCF calculations, however, indicate that the quartet excited state has an energy $\sim$4 eV higher then the doublet; therefore, a spin--orbit-coupling-induced non--radiative intersystem crossing will not be possible. Thus, the SiNV$^0$ represents a well--known two--level quantum system, which is a main building blocks of photonic quantum technologies.

        Upon photoexcitation, the relaxation of the atoms in the SiNV$^0$ is almost unnoticeable; only the Si atom moves slightly towards nitrogen (see Fig.~\ref{fig:Fig7}). The Stokes shift value is therefore very small, only 0.14~eV. The calculated ZPL associated with the a' $\rightarrow$ a'' transition is 0.81~eV, which corresponds to 1530~nm photon emission.
        1530~nm falls within the C band of telecom wavelengths, which is known to exhibit the lowest loss when used in optical transmission systems.
        The SiNV$^0$ center is, therefore, very attractive considering its potential application in scalable quantum telecommunication networks.

        \begin{figure}[t]
        \centering
        \includegraphics[width = 1.0\columnwidth]{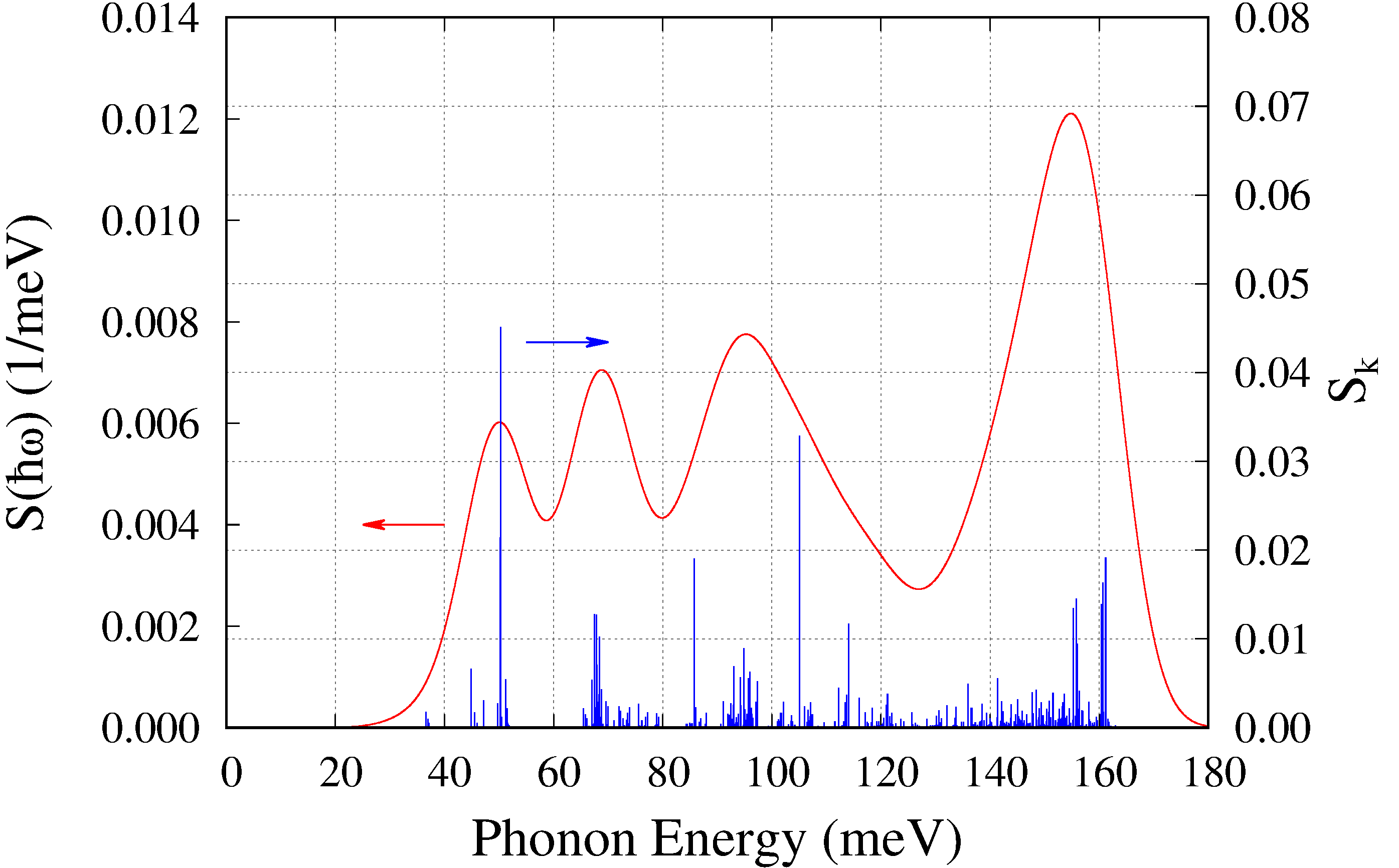}
        \caption{ Partial HR factor (S$_k$) and spectral function S($\hbar{\omega} $) pertaining to the spin-doublet optical transition in the SiNV$^0$ center, as obtained in a 512-atom supercell.}
        \label{fig:Fig8}
        \end{figure}

\begin{figure}[th]
        \centering
        \includegraphics[width=1.0\columnwidth]{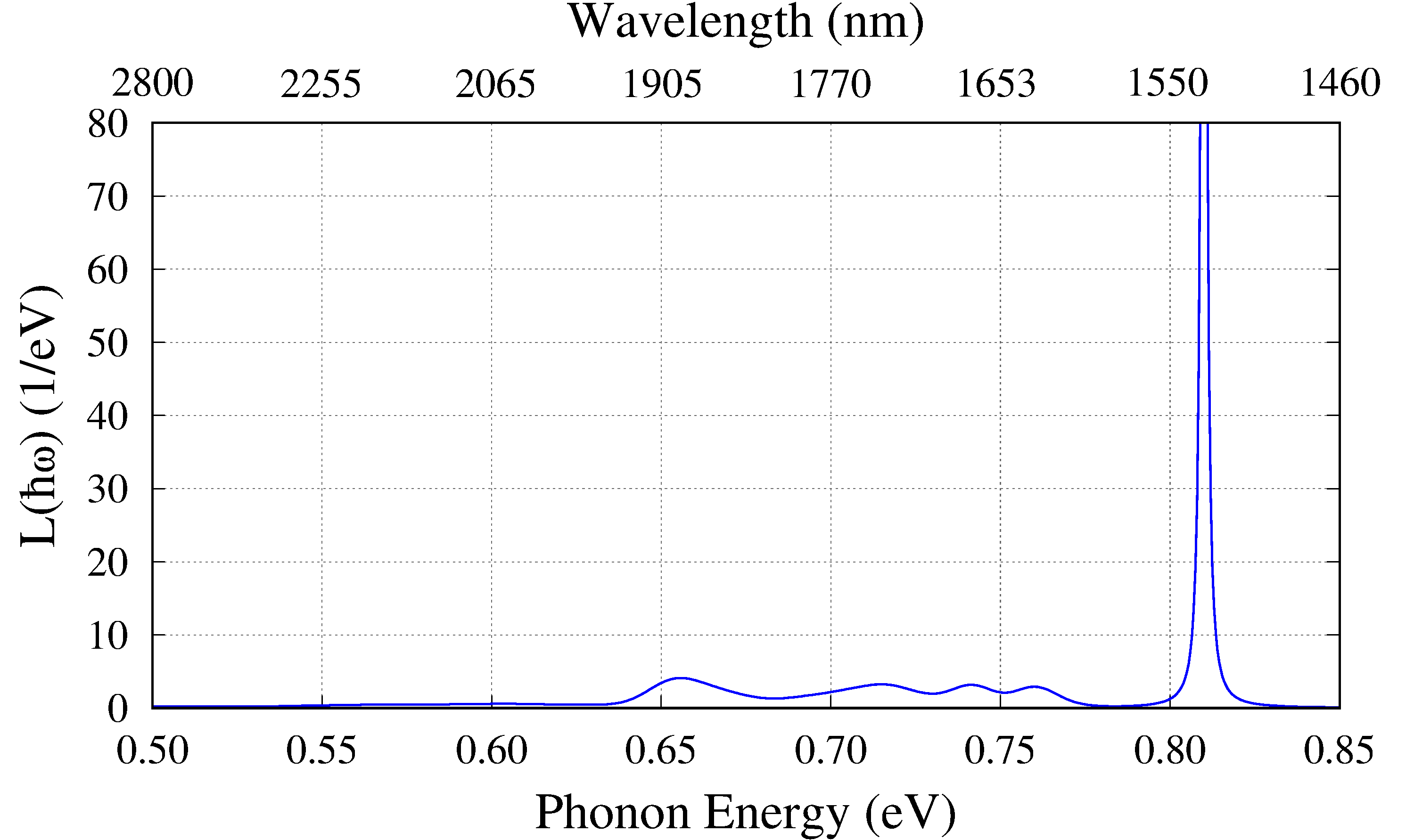}
        \caption{Photoluminescence line shape associated with the lowest energy optical transition in the SiNV$^0$ center.}
        \label{fig:Fig9}
        \end{figure}

        We also simulated the PL spectrum and inspected the HR and DW factors. The partial HR factor S$_k$ together with the spectral function of electron--phonon coupling $S(\hbar\omega)$ are shown in Fig.~\ref{fig:Fig8}. Four weak resonances can be singled out with one clearly associated with the quasi--local vibration at 435 cm$^{-1}$ (54 meV), corresponding to the a'' mode (see Table \ref{tab:modes}).
        Even though the quasi--local mode has theoretically the highest contribution to the phonon sideband ($S_k$=0.045), it constitutes only $\sim$6\% of the total HR factor. We find a total HR factor $S_0$=0.775 and a DW factor 46\%.
        The concentration of intensity in the ZPL is very desirable for quantum applications.  It also highlights the capability of the SiNV$^0$ center to produce single photons. The simulated PL spectrum (see Fig.~\ref{fig:Fig9}) exhibits a sharp peak at the ZPL on top of a weak phonon sideband, which is very similar to the SiV$^-$ center in diamond. Another very important feature of SiNV$^0$ is its low C$_{1h}$ symmetry, which results in an increased dipole transition probability. For this reason, one may expect that emission from SiNV$^0$ will be bright.

        As seen in Fig.~\ref{fig:Fig6}, the Kohn--Sham states involved in the optical transition are relatively close to the valence band, which could be a drawback for quantum applications \cite{Weber2010}.
        We note, however, that the electronic structure of the SiNV$^0$ is similar to that of the well-known SiV$^-$ center \cite{Gali2013PRB}, which is isoelectronic with SiNV$^0$.  The optical transition in the SiV$^-$ center results from exciting an electron out of a Kohn--Sham state that is actually below the VBM in the ground state. Since bright single-photon emission has been demonstrated in the case of SiV$^-$, we believe that the SiNV$^0$ is an equally strong candidate.

        Finally, as mentioned in the Introduction, Wassell \emph{et~al.} speculated that one of the Si$_x$N$_y$ or Si$_x$N$_y$V complexes, where $x$ and $y$ are small integers, might be responsible for the 499~nm line observed in the luminescence spectrum of synthetic CVD diamond. Our calculations clearly exclude the SiN, SiNV, or SiN$_2$V complexes as potential candidates responsible for the emission at 499~nm.

        \section{SUMMARY AND CONCLUSIONS}

       In summary, we used spin-polarized hybrid DFT to investigate the atomic and electronic structure, energetics, and  electrical and magneto-optical properties of various silicon--nitrogen--vacancy complexes in diamond.
        We demonstrated that when Si and N atoms are simultaneously present in the diamond lattice a thermodynamic driving force exists for complexing of these elements, and for the formation of larger complexes with vacancies.
        To thoroughly characterize the complexes, we calculated their hyperfine structure, quasi-local vibrational modes, and optical signatures fully from first principles.
        We provided theoretical evidence that the SiN dimer in diamond acts as a shallow donor with an activation energy comparable to that of substitutional phosphorus.
        We also found that the SiNV$^0$ center is an optically active color center at $\sim$1530 nm, which falls within the C band of telecom wavelengths.
        Our detailed investigation of electron-phonon coupling indicates the promise of the SiNV$^0$ center as a bright single-photon emitter, making it attractive for potential application in scalable quantum telecommunication networks.

    	\begin{acknowledgments}
    		This research was financially supported by the Polish National Science Centre under contract no. UMO-2019/32/C/ST3/00093. Computing resources were provided by High Performance Computing facilities of the Interdisciplinary Centre for Mathematical and Computational Modeling (ICM) of the University of Warsaw under Grant No. GB69-32 and Poznan Supercomputing and Networking Center (PSNC) under Grant No. 275.
            CVdW acknowledges support from the National Science Foundation (NSF) through Enabling Quantum Leap: Convergent Accelerated Discovery Foundries for Quantum Materials Science, Engineering and Information (Q-AMASE-i) award number DMR-1906325.
            JAM acknowledges the support of National Science Centre, under contract No. UMO-2016/23/B/ST3/03567. 

    	\end{acknowledgments}

	%\bibliography{biblio}

\end{document}